\documentclass[12pt]{article}

\usepackage{graphicx}
\usepackage{amsmath}
\usepackage{amssymb}
\usepackage{url}
\usepackage{hyperref}
\usepackage{caption}
\usepackage{listings}

\usepackage[margin=1in]{geometry}

\title{Biometric Bound Credentials for Age Verification}
\author{
  Norman Poh\thanks{Email: npoh@truststamp.net} \\
  Trust Stamp, USA \\
  University of Malta, Malta
  \and Daryl Burns\thanks{Email: dburns@truststamp.net} \\
  Trust Stamp, USA
}
\date{}

\begin{document}

\maketitle

\begin{abstract}
Age verification is increasingly critical for regulatory compliance, user trust, and the protection of minors online. Historically, solutions have struggled with poor accuracy, intrusiveness, and significant security risks. More recently, concerns have shifted toward privacy, surveillance, fairness, and the need for transparent, trustworthy systems. In this paper, we propose Biometric Bound Credentials (BBCreds) as a privacy-preserving approach that cryptographically binds age credentials to an individual’s biometric features without storing biometric templates. This ensures only the legitimate, physically present user can access age-restricted services, prevents credential sharing, and addresses both legacy and emerging challenges in age verification. enhances privacy. 
\end{abstract}


\vspace{0.5em}
\begin{center}
  	{This paper has been peer-reviewed and accepted for presentation at 
    the 24th International Conference of the Biometrics Special Interest Group (BIOSIG 2025)
25.-26.09.2025, Darmstadt, Germany }
\end{center}
\vspace{1em}

\section{Introduction}

Age verification remains a critical challenge in digital environments, particularly for protecting minors from accessing inappropriate content or services. Traditional approaches rely on knowledge factors (passwords, PINs) or possession factors (device ownership, physical ID documents), but are susceptible to vulnerabilities including credential sharing, device access by unauthorized users, and privacy concerns. The most promising solutions to date, e.g., passkeys, still suffer from device theft and device sharing vulnerabilities, as well as unauthorized access by children to their parents' devices, which can lead to a potential `friendly fraud' scenario where minors bypass age restrictions.

Biometric Bound Credentials (BBCreds) offer a promising solution by fundamentally changing how age credentials are secured and accessed. Rather than relying solely on what a user knows or possesses, BBCreds cryptographically bind credentials to the user's biometric characteristics—as an integral part of the authentication process. Unlike conventional biometric authentication systems where biometric verification and credential access are separate processes requiring stored templates, BBCreds create an inseparable cryptographic link between the credential and the user's biometric features without storing actual biometric templates, addressing privacy concerns that have traditionally limited biometric adoption.

This paper explores how BBCreds can revolutionize age verification systems by creating a secure, privacy-preserving mechanism that ensures only the legitimate, physically present user can access age-restricted content or services.

\section{Background and Related Work}

Age verification systems have evolved significantly, driven by regulatory requirements and growing concerns about online safety. Demand has been propelled by the need to fulfill the UK Online Safety Act (2023) and the EU Digital Services Act (2024). In the United Kingdom, initiatives such as the Digital Identity and Attributes Trust Framework \cite{UKDIAF2025} and Authentication Credentials for Online Government Services \cite{GPG44} have established guidelines for secure digital identity verification.

Traditional approaches include document-based verification, knowledge-based verification, credit card verification, mobile network operator verification, and digital identity services. While these methods offer varying levels of assurance, they share common limitations including vulnerability to forgery, information sharing, and assumptions about user age that may not be valid.

Biometric authentication has been proposed as a solution to some of these challenges. The National Cyber Security Centre (UK) provides guidance on biometric recognition technologies for authentication systems \cite{NCSC2025}, acknowledging their potential while highlighting privacy and security considerations. However, traditional biometric systems that store templates create privacy risks and potential security vulnerabilities if compromised.

The innovation of BBCreds lies in their ability to leverage biometric verification without storing actual biometric templates, addressing the privacy concerns that have limited wider adoption of biometric authentication for age verification purposes.

\section{Biometric Bound Credentials: Core Principles}

Biometric Bound Credentials represent a significant advancement in authentication technology by cryptographically binding credentials to an individual's biometric features without storing the actual biometric data. This approach aligns with the ISO/IEC 24745 standard \cite{ISO18013}, which establishes guidelines for biometric information protection.

The core principles that distinguish BBCreds include:

\begin{enumerate}
\item {\bf Person-Centric Authentication}: Unlike device-centric authentication methods that assume the device holder is the authorized user, BBCreds verify the physical presence of the specific individual to whom the credential was issued.

\item{\bf Cryptographic Binding}: BBCreds use cryptographic techniques to bind credentials to biometric features in a way that prevents access without the physical presence of the authorized individual. This binding ensures the credential cannot be extracted without the correct biometric input, the binding process is one-way, and the system can verify authenticity without accessing original biometric data.

\item{\bf Liveness Verification}: A critical component that ensures the biometric sample is being provided by a physically present person rather than a replica or recording, protecting against presentation attacks.

\item{\bf Zero Knowledge Proof Biometrics}: BBCreds implement a zero-knowledge proof approach where authentication occurs without revealing actual biometric data. No party stores biometric templates, the verification process compares cryptographic derivatives rather than actual biometric features, and the system confirms matches without knowing what is being matched.
\end{enumerate}

\section{Technical Implementation for Age Verification}

The implementation of BBCreds for age verification involves a specialized protocol that secures age credentials while ensuring they can only be accessed by the legitimate user. %Figure~\ref{fig:protocol} is a simplified explanation of how this protocol works.

% \begin{figure}[htbp]
% \centering
% \includegraphics[width=0.5\textwidth]{protocol.png}
% \ifdefined\ieee
%   \includegraphics[width=0.5\textwidth]{protocol.png}
% \else
%   \includegraphics[width=1.0\textwidth]{protocol.png}
% \fi
% \caption{Protocol diagram for Biometric Bound Credentials in age verification systems. Bullet 1: Attribute assessment, AgeCred issuance and binding. Bullet 2: BBCred (the encrypted AgeCred) is sent to the user. Bullet 3: Credential unbinding and authentication, releasing the AgeCred that can be used to prove the user's age to a relying party (RP).}
% \label{fig:protocol}
% \end{figure}

\begin{figure*}[htbp]
\centering
\includegraphics[width=\textwidth]{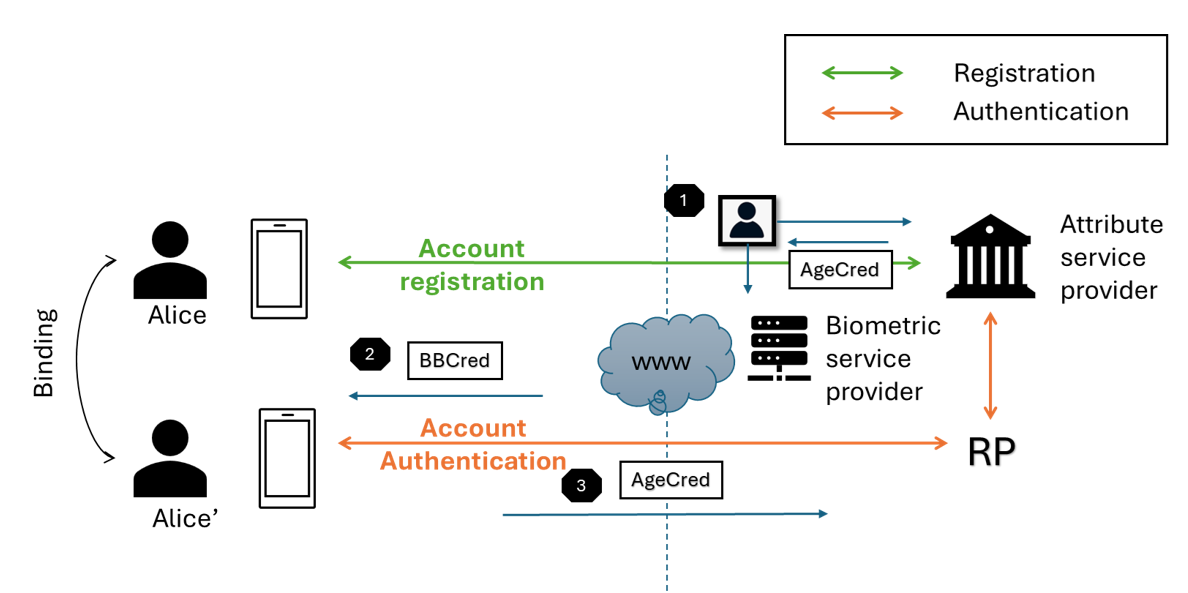}
\caption{Protocol diagram for Biometric Bound Credentials in age verification systems. Bullet 1: Attribute assessment, AgeCred issuance and binding. Bullet 2: BBCred (the encrypted AgeCred) is sent to the user. Bullet 3: Credential unbinding and authentication, releasing the AgeCred that can be used to prove the user's age to a relying party (RP).}
\label{fig:protocol}
\end{figure*}

We shall use the term ``system'' to be context independent. Typically, the capture system is a mobile device that has a camera and can perform biometric capture and processing on the device. This processing includes liveness detection, biometric feature extraction, and the cryptographic operations needed for credential binding and unbinding. All secrets are processed locally on the user's device, and no biometric data is transmitted to any remote server. 

During the initial enrollment phase, the user's age must be verified, which may require the device to transmit some personal information to the attribute service provider (ASP). The ASP can verify the user's age through appropriate channels, e.g., age estimation from a selfie capture, document verification, digital identity services, or mobile network operator checks \cite{Ofcom2025}. This sharing of personal information is limited to the enrollment phase only; post-issuance of the BBCred, no sensitive data needs to be shared during subsequent authentication processes.

The binding of biometric data to credentials in this system is accomplished using a fuzzy extractor, as described by Dodis et al. \cite{Dodis2004}, and more recent neural fuzzy extractor approaches \cite{Jana2020}. These cryptographic techniques allow for the derivation of stable keys from inherently noisy biometric data, enabling consistent authentication while accommodating natural variations in biometric samples.

In our implementation, we use a proprietary fuzzy extractor that is significantly more advanced and secure. It is capable of extracting 256 bits of entropy from a single biometric reference with 512 dimensions, ensuring compliance with NIST SP 800-63B guidelines for cryptographic strength in authentication systems. This proprietary method leverages advanced cancellable biometric representation, error correction, and feature selection techniques. As a result, it allows reliable key regeneration from high-dimensional biometric vectors (e.g., 512 dimensions) while maintaining robustness against intra-user variability and attacks.

Our fuzzy extractor improves upon the method by Jana et al. \cite{Jana2020}, which can extract only 128 bits of entropy instead of 256. However, for the central thesis and contribution of this paper, the specific choice of fuzzy extractor is immaterial as long as it can reliably extract a stable key from the biometric data, with sufficient key length to meet the security requirement of being post-quantum ready.

\subsection{Issuance of AgeCred and BBCred (Enrollment Phase)}
The process begins with the issuance of an age credential (AgeCred) that is then bound to the user's biometric features:
\begin{enumerate}
\item The user undergoes a biometric capture session, e.g., taking a selfie.
\item The system performs liveness detection to verify that the sample comes from a physically present person.
\item The biometric sample is processed using a fuzzy extractor \cite{Dodis2004}, generating a stable cryptographic key (StableKey) that can be consistently reproduced from future samples of the same person's biometrics.
\begin{equation}
  \text{StableKey} = \text{FuzzyExtract}(\text{Embedding}_{enrol}, \text{HelperData})
\end{equation}
\item The hash of the StableKey is stored securely, 
\begin{equation}
  \text{HashStableKey} = \text{Hash}(\text{StableKey}),
\end{equation}

\item The system generates a Sketch, which assists in reproducing the StableKey from future biometric samples:
   \begin{equation}
   \text{StableSecret} = \text{StableKey} \oplus \text{Sketch}
   \end{equation}
    where $\oplus$ denotes the XOR operation.
    The Sketch protects both the StableKey and StableSecret. An attacker who gains access to the Sketch cannot derive either without the original biometric sample.
\item An attribute service provider verifies the user's age through appropriate channels.
\item Upon successful age verification, an age credential (AgeCred) is issued.
\item The AgeCred is cryptographically bound to the biometric data via encryption:
  \begin{equation}
    \text{BBCred} = \text{Encrypt}[\text{StableSecret}](\text{AgeCred})
  \end{equation}
  where Encrypt is a symmetric encryption function using the StableSecret as the key.
\item Since the StableKey is no longer needed, it is discarded immediately.
\item The resulting BBCred is stored on the user's device.
\end{enumerate}

At the end of enrollment, the system only retains the Sketch, the encrypted AgeCred (BBCred), the hash of the StableKey (HashStableKey), and the HelperData. None of these components can be used to reconstruct the original biometric sample or the StableKey, ensuring privacy and security.

\subsection{Authentication and Credential Unbinding}
When the user attempts to access age-restricted content or services:
\begin{enumerate}
\item The user provides a new biometric sample.
\item Liveness detection verifies the physical presence of the user.
\item The system processes the biometric sample to extract a new StableKey$'$.
  \begin{equation}
  \text{StableKey}' = \text{FuzzyExtract}(\text{Embedding}_{auth}, \text{HelperData})
  \end{equation}
  If the hash of StableKey$'$ doesn't match the stored hash, authentication fails.
\item Using the stored Sketch, the system attempts to reproduce the original StableSecret:
  \begin{equation}
  \text{StableSecret}' = \text{StableKey}' \oplus \text{Sketch}
  \end{equation}
  If StableKey$'$ is derived from a different person's biometric sample, the resulting StableSecret$'$ will not match the original StableSecret, preventing unauthorized access.
\item The system attempts to decrypt the BBCred using the reproduced StableSecret$'$:
  \begin{equation}
  \text{AgeCred} = \text{Decrypt}[\text{StableSecret}'](\text{BBCred})
  \end{equation}
  If decryption succeeds, the AgeCred is considered valid, and access is granted.
\end{enumerate}

This process ensures that only the registered user, physically present at authentication time, can access the age credential.

\subsection{Protocol Variations}
There are several possible variations in this protocol: 
\begin{enumerate}
  \item {\bf One-way function for binding}: When the AgeCred is generated internally, it can be bound to the StableSecret using a one-way function:
    \[\text{AgeCred} = \text{Hash}(\text{StableSecret} \oplus \text{Sketch})\]
    This method prevents reversal attacks, as compromising the AgeCred does not allow reconstruction of the StableSecret.

  \item {\bf Sketch protection via symmetric encryption}: Instead of using XOR, the Sketch can be created by encrypting the StableSecret with the StableKey:
    \[\text{Sketch} = \text{Encrypt}[\text{StableKey}](\text{StableSecret})\]
    Symmetric encryption offers stronger protection than XOR, as it prevents an attacker from recovering the StableSecret or StableKey even if the Sketch is compromised.
\end{enumerate}
\section{Advantages for Age Verification Applications}
BBCreds offer several significant advantages for age verification applications:

\begin{enumerate}
\item{\bf Prevention of Credential Sharing}: One of the most significant advantages of BBCreds in age verification is the prevention of credential sharing. Unlike passwords or PINs that can be easily shared, biometric binding ensures that only the registered individual can use the credential. This directly addresses the `friendly fraud' scenario where children might use their parents' or older siblings' devices to access age-restricted content.

\item{\bf Enhanced Privacy Protection}: BBCreds protect user privacy in several ways. The system eliminates storage of actual biometric templates, significantly reducing the risk of biometric data breaches. It adheres to data minimization principles through minimal data collection and storage requirements. The verification process is decentralized and can occur on the user's device without transmitting biometric data to external servers. Additionally, the BBCred approach is designed for compliance with privacy regulations like GDPR from the ground up, incorporating privacy by design principles throughout the system architecture.

\item{\bf Improved User Experience}: While enhancing security, BBCreds also improve the user experience. Users benefit from seamless authentication without needing to remember complex passwords or PINs. The system offers reduced friction compared to document-based verification methods that require manual uploads and verification steps. Furthermore, BBCreds have the potential for reusable verification across multiple services, eliminating the need for repetitive verification processes and creating a more streamlined user journey across different platforms and applications.

\item{\bf Resistance to Common Attack Vectors}: BBCreds can resist common attack vectors that plague traditional authentication systems. Phishing attacks become ineffective as there are no credentials for attackers to steal through deceptive means. Credential stuffing and brute force attacks are not applicable to this biometric-bound approach. Man-in-the-middle attacks cannot capture usable authentication data due to the local processing of biometric information. Perhaps most importantly, device theft does not compromise the age verification system, as the physical presence of the registered user is required for successful authentication, creating a fundamentally more secure verification mechanism.

\item{\bf Regulatory Compliance}: BBCreds enable service providers to meet stringent regulatory requirements for age verification through a balanced approach to security and privacy. By providing strong assurance against underage access while maintaining audit trails without storing sensitive personal data, the system simultaneously addresses compliance mandates and privacy concerns. This proportionate solution aligns with modern regulatory frameworks that demand both effective age verification and robust data protection.
\end{enumerate}

\section{Limitations and Future Directions}

Despite their advantages, BBCreds face several limitations and challenges. False acceptance due to similar-looking individuals presents a challenge, particularly with identical twins. This can be addressed by incorporating additional biometric modalities like iris or fingerprint recognition. False rejection in adverse capture environments is another concern, as poor lighting or suboptimal camera positioning can lead to failed authentication attempts. This can be mitigated through active user interface guidance during capture. Biometric changes over time also pose challenges, as significant appearance changes can cause recognition failures. Implementing adaptive biometric systems \cite{Pisani2019} that update enrollment samples after successful authentications can address this issue.

User acceptance remains a significant barrier, as some users may hesitate to use biometric authentication due to privacy concerns. Education about the privacy-preserving nature of BBCreds will be essential for broader adoption. Accessibility considerations must be addressed to accommodate users with disabilities or those whose biometric characteristics may be difficult to capture reliably. Alternative authentication methods must be provided for users who cannot or choose not to use biometric verification, creating a more inclusive verification ecosystem.

The evolution of BBCreds for age verification presents several promising research and development directions: Multi-modal Biometric Binding could enhance security and reliability by incorporating multiple biometric modalities while maintaining privacy-preserving properties. Federated Age Verification would allow users to verify their age once and use that verification across multiple services, improving user experience while maintaining security. Adaptive Systems research is crucial for accommodating gradual changes in biometric characteristics over time, ensuring long-term effectiveness despite natural aging processes. Standardization Efforts will play a vital role in promoting interoperability, establishing best practices, and facilitating wider adoption across different platforms and services.

\section{Conclusion}

Biometric Bound Credentials represent a transformative approach to age verification in digital environments. By shifting the trust anchor from the device to the person, BBCreds address fundamental limitations in current age verification systems, particularly the vulnerability to credential sharing and device access by unauthorized users.

The privacy-preserving nature of BBCreds, which operate without storing actual biometric templates, offers a compelling solution to privacy concerns that have limited biometric authentication adoption. By implementing zero-knowledge proof biometrics in accordance with standards like ISO/IEC 24745 \cite{ISO18013}, BBCreds provide strong security guarantees while respecting user privacy.

While challenges remain—particularly related to biometric accuracy, implementation complexity, and user adoption—the advantages of BBCreds for age verification are significant. The ability to prevent credential sharing, enhance privacy protection, improve user experience, and resist common attack vectors makes BBCreds a promising technology for meeting growing regulatory and social demands for effective age verification.

As digital services continue to evolve and age verification requirements become more stringent, BBCreds offer a path toward more secure, privacy-respecting, and user-friendly age verification systems. By reimagining digital trust through biometric binding, we can create age verification mechanisms that truly verify the presence of the authorized individual, rather than merely checking for knowledge of a password or possession of a device.

\bibliographystyle{plain}
\bibliography{references.bib}

\end{document}